\documentclass[twocolumn]{aastex631}

\usepackage{physics}
\usepackage{physunits}
\usepackage{upgreek}

\shorttitle{Hadronic vs. Leptonic}
\shortauthors{Corso}

\begin{document}

\title{Hadronic versus leptonic origin of gamma-ray emission from supernova remnants}

\author[0000-0002-0088-2563]{Nicholas J. Corso}
\affiliation{Department of Astronomy and Astrophysics, The University of Chicago, 5640 S Ellis Ave, Chicago, IL 60637, USA}
\affiliation{Department of Astronomy, Cornell University, 404 Space Sciences Building, Ithaca, NY 14853, USA}

\author[0000-0002-6679-0012]{Rebecca Diesing}
\affiliation{Department of Astronomy and Astrophysics, The University of Chicago, 5640 S Ellis Ave, Chicago, IL 60637, USA}

\author[0000-0003-0939-8775]{Damiano Caprioli}
\affiliation{Department of Astronomy and Astrophysics, The University of Chicago, 5640 S Ellis Ave, Chicago, IL 60637, USA}
\affiliation{Enrico Fermi Institute, The University of Chicago, Chicago, IL 60637, USA}

\begin{abstract}

GeV and TeV emission from the forward shocks of supernova remnants (SNRs) indicates that they are capable particle accelerators, making them promising sources of Galactic cosmic rays (CRs).
However, it remains uncertain whether this $\gamma$-ray emission arises primarily from the decay of neutral pions produced by very high energy hadrons, or from inverse-Compton and/or bremsstrahlung emission from relativistic leptons.
By applying a semi-analytic approach to non-linear diffusive shock acceleration (NLDSA) and calculating the particle and photon spectra produced in different astrophysical environments, we parametrize the relative strength of hadronic and leptonic emission.
We show that, even if CR acceleration is likely to occur in all SNRs,  the observed photon spectra may instead primarily reflect the environment surrounding the SNR, specifically the ambient density and radiation field.
We find that the most hadronic-appearing spectra are young and found in environments of high density but low radiation energy density.
This study aims to guide the interpretation of current $\gamma$-ray observations and single out the best targets of future campaigns.

\end{abstract}

\section{Introduction} \label{sec:intro}

The forward shocks of supernova remnants (SNRs) are promising candidates for the primary sources of Galactic cosmic rays (CRs) since they provide sufficient energetics and an efficient acceleration mechanism, diffusive shock acceleration, or DSA \citep{drury+94, hillas05, berezhko+07,ptuskin+10,caprioli+10a}. 
However, direct evidence of efficient hadron acceleration by SNRs, particularly up to the so-called CR knee at energies $\gtrsim 10^{15}$ eV, remains limited \citep{blasi19}.

The best observational evidence for hadron acceleration is $>100\MeV$ $\gamma$-ray emission from the decay of neutral pions ($\pi_0$) produced by interactions between CR ions and the ambient medium \citep{drury+94}. 
However, when leptons---primarily electrons---are accelerated, they, too can produce strong $\gamma$-ray signatures via inverse Compton (IC) and relativistic bremsstrahlung radiation \citep{aharonian+06}. 
Ideally, observational signatures would identify which of the two scenarios dominates, but in many cases the results are ambiguous. 

One example of this ambiguity is RX J1713.7-3946, which was identified as a source of high-energy $\gamma$-ray emission when it was detected in the TeV-band by the HESS collaboration \citep{aharonian+06}. 
As demonstrated in \cite{morlino+09}, both hadronic and leptonic scenarios could explain observed HESS data. 
Later data from Fermi-LAT, however, favored leptonic models, apparently indicating that RX J1713.7-3946 and Vela Jr. are not efficient hadron accelerators \citep{ellison+10,zirakashvili+10,abdo+11,lee+13}.
On the other hand, \cite{morlino+12} later identified Tycho's SNR as a strong candidate for hadronic $\gamma$-ray based on VERITAS and Fermi-LAT data;
also, the detection of the characteristic ``pion bump" in IC443 and W44 confirmed the hadronic nature of the $\gamma$-ray emission from these SNRs interacting with molecular clouds \citep[][]{ackermann+13}. 
These findings raise questions about whether hadronic emission from SNRs is common enough for them to be the primary accelerators of Galactic CRs.

Ideally, to assess the hadronic/leptonic nature of a source, multi-wavelength measurements would be made for all $\gamma$-ray bright SNRs in consideration, but this is time-consuming and often inconclusive due to the often limited constraints on age and distance.

In the recent years, kinetic simulations of non-relativistic shocks have shown that the  acceleration of protons and heavier nuclei is mostly controlled by the local inclination of the shock (the angle between the shock normal and the local magnetic field), rather than by the shock strength, parametrized by its sonic and Alfv\'enic Mach numbers \citep[][]{caprioli+14a,caprioli+14b,caprioli+14c,caprioli+17,caprioli+18,haggerty+20}.
Conversely, the efficiency of electron acceleration in such shocks is not fully understood, yet \citep[though see the works by][]{guo+14a,park+15,xu+20,shalaby+22}.

Since most SNRs are likely probing a variety of shock inclinations \citep[][]{pais+20,winner+20}, we do not expect the acceleration efficiency to vary greatly in all of the SNRs that exhibit strong shocks.
Therefore, we propose that environmental factors are essential determiners of the dominant emission mechanism. 
That is, we consider how the age and the characteristics of the medium in which SNRs expand (density profile and normalization, energy density of background radiation), impact their $\gamma$-ray production and thus its inferred hadronic/leptonic nature.


In this study, we apply a semi-analytic formalism for non-linear diffusive shock acceleration to calculate the particle and photon spectra at various stages in the evolution of simulated SNRs. 
Using these spectra, we parameterize the relative strength of hadronic and leptonic emission in order to assess when the former dominates over the latter. 

Such an analysis, which leverages on the relative normalization and spectral slopes of different kinds of $\gamma$-ray emission, provides information about when/where SNRs are most likely to exhibit a hadronic signature.
We construct effective ``look-up tables" that may guide the interpretation of the spectra currently available and of those that will be provided by the incoming generation of $\gamma$-ray telescopes, such as LHAASO and especially CTA.

The paper is organized as follows: in Section \ref{sec:methods}, we discuss the theoretical background of our computational model, along with the analytical tools with which we use to present our results; and in Sections \ref{sec:results} and \ref{sec:discuss}, we present and analyze our results, with the goal of providing a tool for quick estimation of a source's capacity to produce hadronic emission.

\section{Method} \label{sec:methods}

In general, the evolution of a SNR follows four principal stages \citep[e.g.,][]{bisnovatyi-kogan+95,ostriker+88,diesing+18}: 
in the \textit{ejecta-dominated stage}, the ejected mass is much greater than swept-up mass and the SNR expands effectively unimpeded; 
in the \textit{Sedov stage}, the swept-up mass exceeds the ejected mass and the SNR expands adiabatically; 
in the \textit{pressure-driven snowplow}, the SNR begins to lose energy to radiative cooling but continues to expand due to its internal pressure exceeding that of the ambient background; 
finally, in the \textit{momentum-driven snowplow}, expansion is driven by the residual kinetic energy from the explosion. 
For this work, only the ejecta-dominated and Sedov stages are considered, since most of the SNRs seem to fade away in the radiative stage \citep[e.g.,][]{case-bhattacharia98,bandiera+10}.

Broadly speaking, we consider two types of environments surrounding our model SNRs. In the first, we assume a homogeneous interstellar medium (ISM) with a uniform matter density. 
In the second, we model an environment that may exist around a core-collapse supernova, in which the medium is dominated by stellar winds driven by the supernova progenitor and exhibits an inverse square matter density profile ($n \propto R_{\rm sh}^{-2}$).

All simulated SNRs eject $1M_{\odot}$ of mass with $E=10^{51}\erg$ of kinetic energy. 
For the homogeneous profiles, we assume an ambient magnetic field strength of $B_0=3\Gauss[\upmu]$, and we test ambient number density values spanning $n_0\in[5\times10^{-3},10^1]\cm^{-3}$. 
In the case of the wind profile, we represent the number density as $n(r)=n_0\left(r/\pc\right)^{-2}$.
Our choice of $n_0$ is motivated by \cite{weaver+77}, who note that $\rho(r)\propto \dot{M}/(V_w r^2)$, where $\dot{M}$ is the mass loss rate of the progenitor and $V_w$ is the speed of its stellar wind. Taking $\dot{M}_{-5,\odot}/V_{w,6}$ to vary around order unity, with $\dot{M}=\dot{M}_{-5,\odot} 10^{-5} M_{\odot} \y^{-1}$ and $V_w = V_{w,6} 10^6 \km \s^{-1}$, we obtain and sample the following range of values: $n_0\in[3.5\times10^{-2},1.1\times10^1]\cm^{-3}$.
We adopt an ambient magnetic field strength profile that goes as the square root of the number density, of the form $B_{0} / \Gauss \simeq 0.01 \sqrt{n /\left(5000 \cm^{-3}\right)}$ \citep{chevalier98}.

We simulate CR acceleration using the semi-analytic formalism for non-linear diffusive shock acceleration (NLDSA) described by \cite{caprioli+09a}, \cite{caprioli+10b}, \cite{caprioli12}, and  \cite{diesing+19}, and references therein \citep{malkov97,malkov+00,blasi02,blasi04,amato+05,amato+06}. 
This model self-consistently solves the diffusion-advection equation for the transport of non-thermal particles in a quasi-parallel, non-relativistic shock, including the dynamical backreaction of accelerated particles and of CR-generated magnetic turbulence. 
Magnetic field amplification due to CR-driven streaming instabilities is taken into account as described in \cite{diesing+21};
more precisely, fast shocks are dominated by the non-resonant (Bell) instability, while later in the Sedov stage the resonant instability becomes important \citep[e.g.,][]{bell04,amato+09}.

\begin{figure*}[ht]
\begin{center}
\includegraphics[width=0.95\textwidth, clip=true,trim= 10 10 10 20]{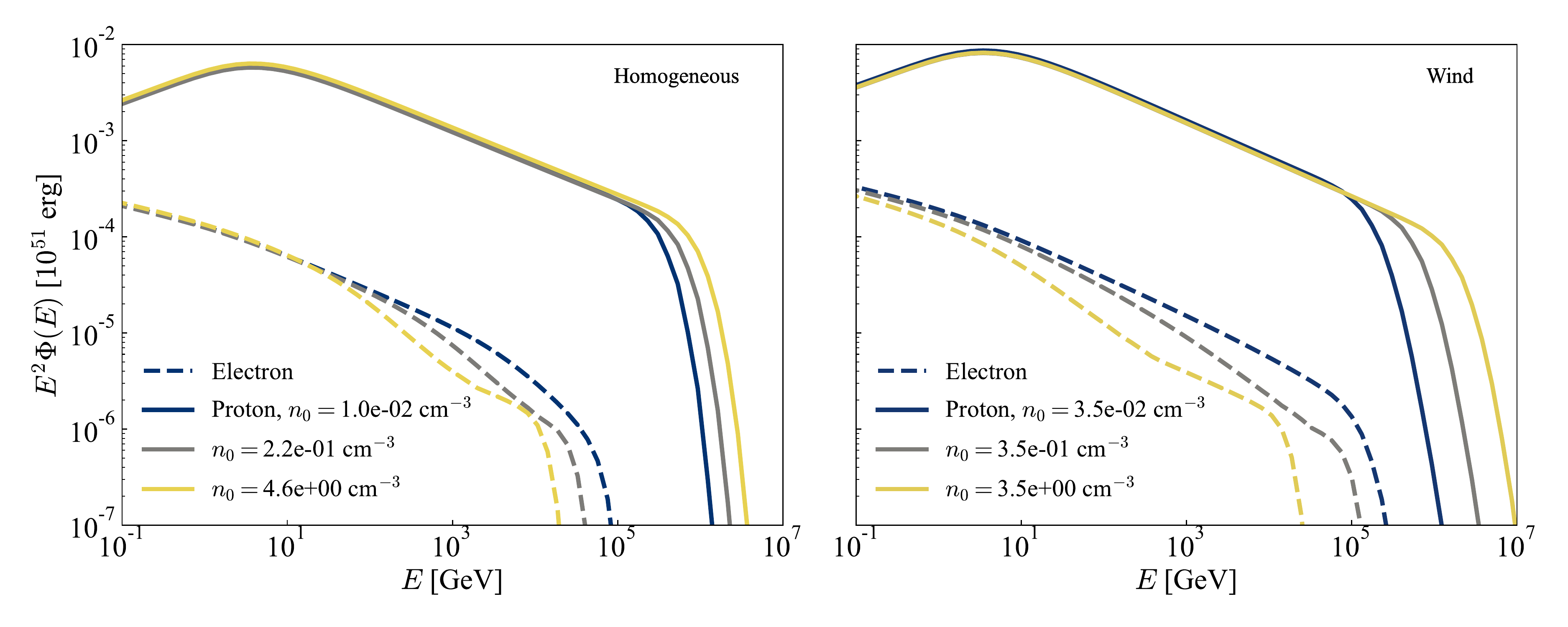}
\caption{Cumulative proton spectra (solid lines) and electron spectra (dashed lines) at the transition between the ejecta-dominated and Sedov-Taylor stages of a modeled SNR in a sample of environments. 
Line colors denote the density normalization used in each model. In the left panel, the ambient medium is taken to be homogeneous; in the right, it follows a wind profile. \label{particle_spectra}}
\end{center}
\end{figure*}

This formalism calculates the instantaneous proton spectrum at each timestep of SNR evolution. As in \cite{diesing+19}, we calculate the instantaneous electron spectra using the analytical approximation provided in \cite{zirakashvili+07},
\begin{equation}
f_{\mathrm{e}} (p)=K_{\mathrm{ep}} f_{\mathrm{p}}(p)\left[1+0.523\left(p / p_{\mathrm{e}, \max }\right)^{9 / 4}\right]^{2} e^{-p^{2} / p_{\mathrm{e}, \max }^{2}},
\end{equation}
with $p_{\mathrm{e}, \max }$ the maximum electron momentum determined by equating the acceleration and synchrotron loss timescales and $K_{\mathrm{ep}}$ the normalization of the electron spectrum relative to that of protons. 
Our reference value is $K_{\mathrm{ep}}=1.6\times10^{-3}$, which corresponds to the value determined for Tycho's SNR in \cite{morlino+12}, and discuss how results change when varying this parameter over the range of $K_{\mathrm{ep}}=10^{-4}$--$10^{-2}$, which encompasses the range of values inferred in other SNRs as well \citep{berezhko+04a, berezhko+06b, berezhko+06a, lee+13}. 
The instantaneous proton and electron spectra are then weighted to account for adiabatic and--in the case of electrons--synchrotron losses, before being summed to produce a cumulative spectrum \citep[see][for more details]{caprioli+10a, diesing+19}. 

To generate the spectrum of nonthermal radiation from a modeled SNR, we use the {\tt Naima} Python package \citep{naima} which, given arbitrary proton and electron momentum distributions, calculates the emission due to IC \citep{khangulyan+14}, synchrotron \citep{aharonian+10}, nonthermal bremsstrahlung \citep{baring+99}, and pion decay \citep{kafexhiu+14}.

We consider different background radiation fields, meant to mimic different astrophysical environments, on top of the ubiquitous Cosmic Microwave Background radiation (CMB), with a temperature of $T=2.72\K$ and an energy density $u_{\rm rad}=0.261\eV/\cm^3$. 
An effective ``maximal" radiation field would correspond to that of a HII environment, as described in Section 12.7 of \cite{draine11}, with an energy density of $u_{\rm rad}=3.9\times10^3\eV/\cm^3$. 
To span the range of photon energy densities between these two extremes, we also consider an environment consisting of the CMB field and starlight peaking in the mid-infrared (MIR) with a temperature of $T=100\K$.
We treat the energy density of this stellar radiation field as a free parameter.

To interpret the dominant form of emission, we introduce a parameter $H$, which we name the \textit{hadronicity} of the emitted radiation. 
This parameter is defined as:
\begin{equation}
H\equiv\frac{2}{\pi}\arctan 
\left[\log_{10}\left( \frac{L_{\text{had}}}{L_{\text{lep}}}\right)\right]
 \label{Hadronicity}
\end{equation}
where $L_{\rm had/lep}$ is the hadronic/leptonic luminosity integrated over a given energy band.
Throughout this work, we consider the ``GeV'' band as $100\MeV$--$100\eV[G]$ and the ``TeV'' band as $100 \eV[G]$--$1\eV[P]$. 
These two bands broadly reflect the regimes of energy spanned by GeV and TeV observatories (i.e., Fermi and Cherenkov telescopes). 
Using this definition, a value of $0.75<H\leq1.$ is considered ``extremely hadronic.'', while $0.<H\leq0.5$ corresponds to ``mildly hadronic.'' 
Conversely, similar absolute values of $H$ but with negative sign would correspond to extremely and mildly leptonic cases.
The hadronicity parameter $H$ may be interpreted as the likelihood that the $\gamma$-ray emission of SNR with given characteristics (age, density, magnetic field, photon background) is of hadronic origin. 

\begin{figure*}[ht]
\begin{center}
\includegraphics[width=0.95\textwidth, clip=true,trim= 10 10 10 15]{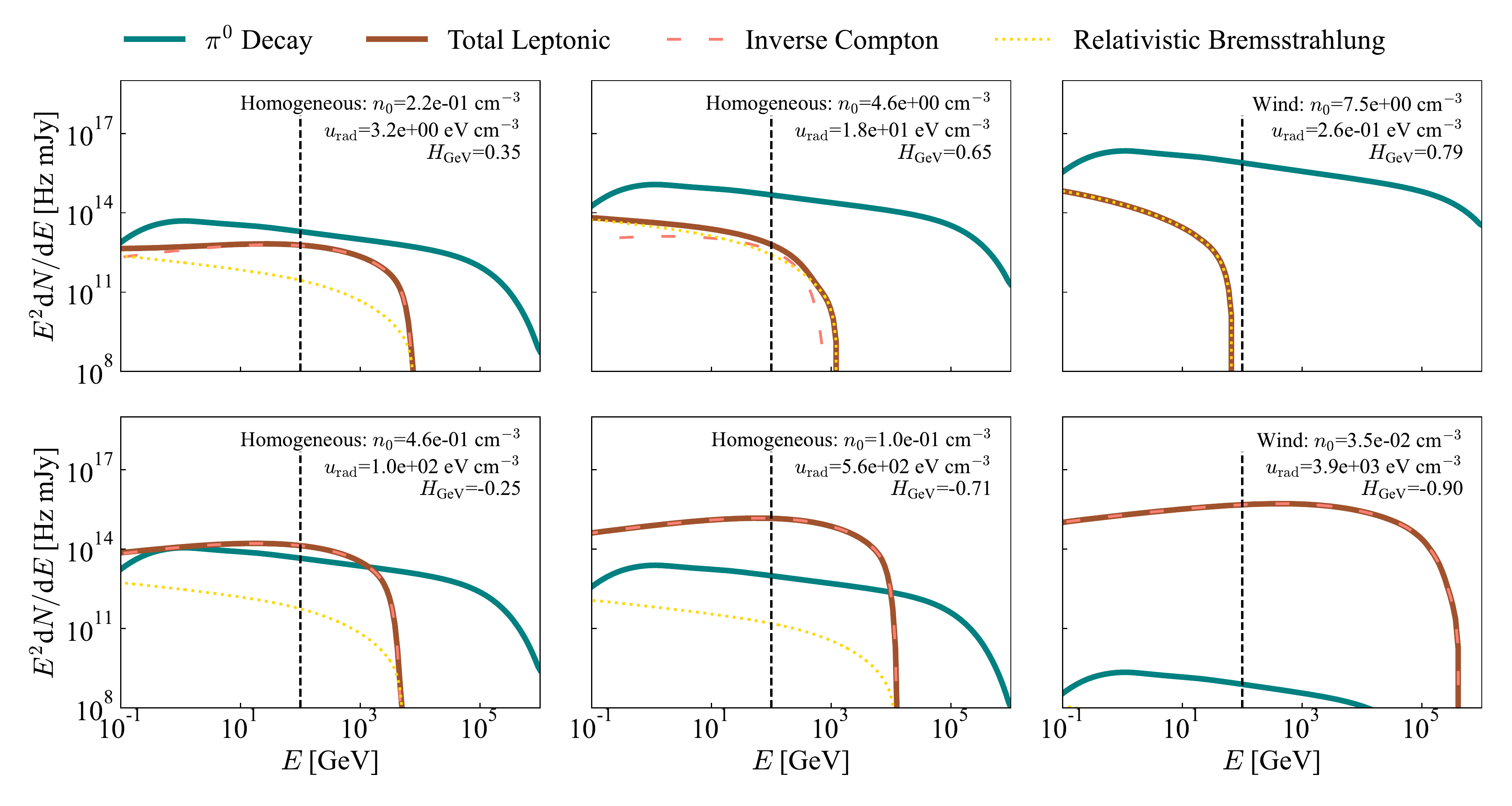}
\caption{Cumulative hadronic  $\gamma$-ray  spectra (blue lines) and leptonic $\gamma$-ray spectra (red lines) at the transition between the ejecta-dominated and Sedov-Taylor stages of a modeled SNR in a sample of environments. Each panel represents a different ambient medium, with $n_0$ denoting the matter density normalization and $u_{\rm rad}$ denoting the radiation energy density. As in Figure \ref{particle_spectra}, we consider both homogeneous and wind profiles for the ambient density. Vertical black dashed lines mark $100\eV[G]$, the dividing energy between our GeV and TeV bands. The top (bottom) panels represent hadronic (leptonic) cases with the extremity of the scenario (i.e., the absolute value of the hadronicity, $H$) increasing from left to right. Note that the nature of the underlying particle acceleration remains the same across panels; the strong variation in $\gamma$-ray emission shown here arises solely from environmental factors \label{photon_spectra}}
\end{center}
\end{figure*}

For practical purposes, $H$ is meant to provide an informed guess about the hadronic/leptonic nature of the GeV or TeV emission from a given SNR without the need of performing a detailed time-dependent, multi-zone, calculation of particle acceleration and its ensuing  multi-wavelength emission.

\section{Results} \label{sec:results}

The ion and electron spectra produced by a sample of SNRs expanding in different density profiles are shown in Figure \ref{particle_spectra}. 
These spectra are the comulative post-shock distributions calculated when the SNR transitions from the ejecta-dominated to the Sedov stage. This time, denoted $T_{\mathrm{ST}}$, is evaluated to be the moment that the accumulated mass from the surrounding medium exceeds the originally ejected mass.

It is worth stressing that these spectra are steeper than the standard DSA prediction, $dN/dE \propto E^{-2}$, due to the shock modification induced by nonthermal particles and by the amplified magnetic they generate \citep[in particular, we include the effects of a ``postcursor," as described in ][]{haggerty+20, caprioli+20,diesing+21}. 
In addition, electron spectra are cooled by inverse-Compton and synchrotron losses, as pointed out by \citet{diesing+19} and discussed by \citet{cristofari+21, morlino+21};
since the maximum electron energy is controlled by synchrotron losses, it is strongly dependent on the amplified magnetic field, which correlates with the local density, too. 

\begin{figure*}[ht]
\includegraphics[width=0.95\textwidth, clip=true,trim= 10 10 10 15]{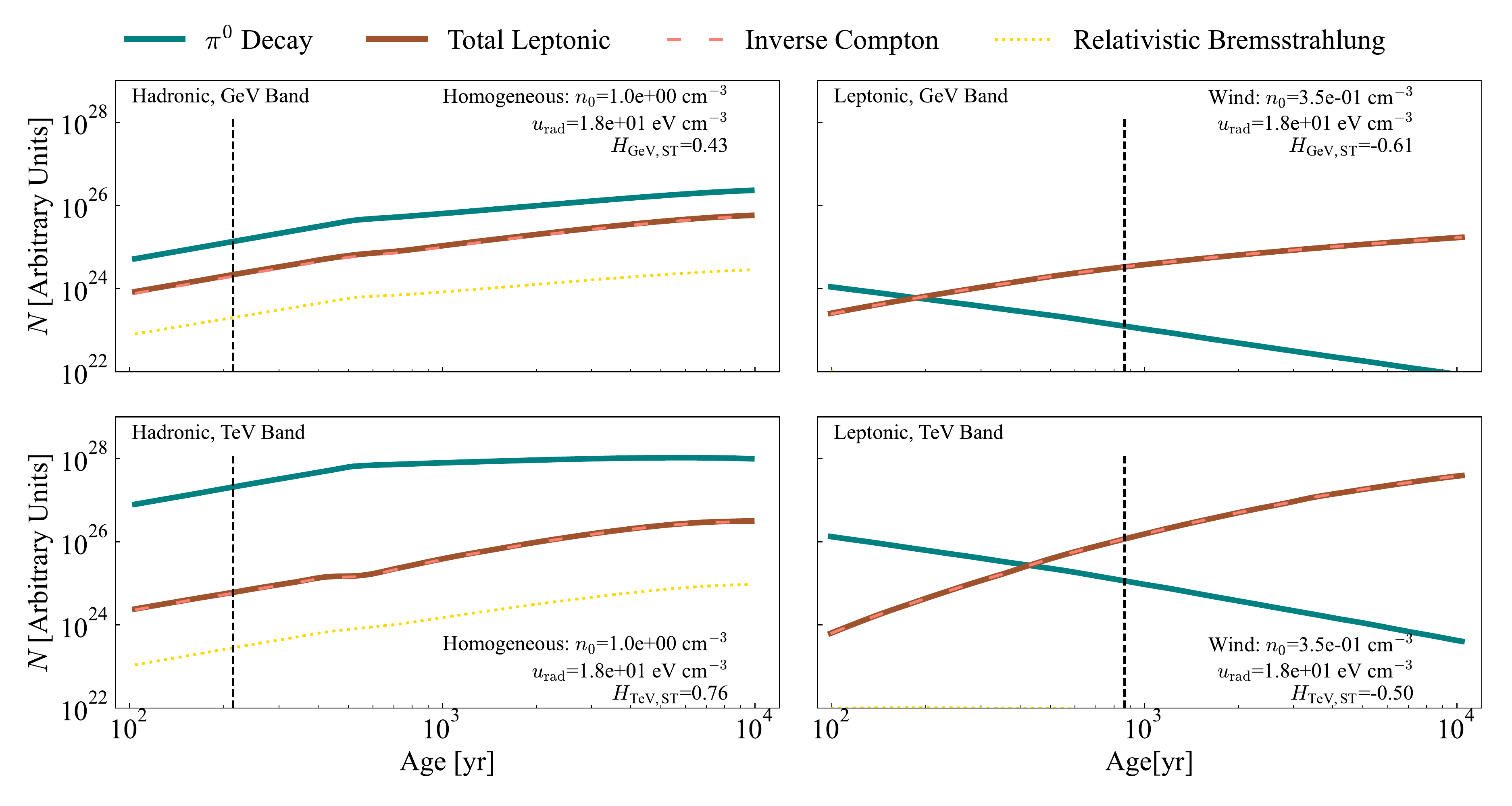}
\caption{Time evolution of hadronic and leptonic luminosities for a moderately hadronic homogeneous profile scenario (left) and a moderately leptonic wind profile scenario (right). Environmental parameters are the same as those used in the middle column of Figure \ref{photon_spectra}.  The top (bottom) row corresponds to luminosities calculated in the GeV (TeV) band. The vertical black dashed lines denote onset of the Sedov-Taylor phase. 
\label{lum_time}}
\end{figure*}

In Figure \ref{photon_spectra}, we present a wide sample of modeled $\gamma$-ray spectra from SNRs in different environments, spanning the full range of hadronicity.
Also these spectra are calculated at the beginning of the Sedov stage.
This variation is exacerbated by the fact that an increase in the hadronic $\gamma$-ray luminosity typically accompanies a decrease in the leptonic luminosity and vice-versa. 
Namely, the luminosity due to $\pi_0$-decay scales the ambient density, while IC emission tends to be inhibited in denser environments, where electrons tend to suffer strong synchrotron losses. 
Note also that relativistic bremsstrahlung is always subdominant with respect to IC.

Figure \ref{lum_time}, instead, depicts the time evolution of the GeV and TeV luminosities for a homogeneous profile and a wind one.
In particular, for the homogeneous scenario, in the late Sedov stage the leptonic luminosity tends to grow faster than the hadronic luminosity, as a consequence of the shifting of the electron cut-off to higher energies when the amplified magnetic field decreases and synchrotron losses become less severe. As a consequence, over time, there is a general trend for the spectra to become more leptonic, perhaps even switching from being dominantly hadronic to leptonic. In the case of a wind profile, the leptonic emission monotonically increases against a monotonically decreasing hadronic curve, which results in a definitive progression toward a leptonically dominated scenario.

\begin{figure*}[ht]
\includegraphics[width=0.95\textwidth, clip=true,trim= 10 10 10 20]{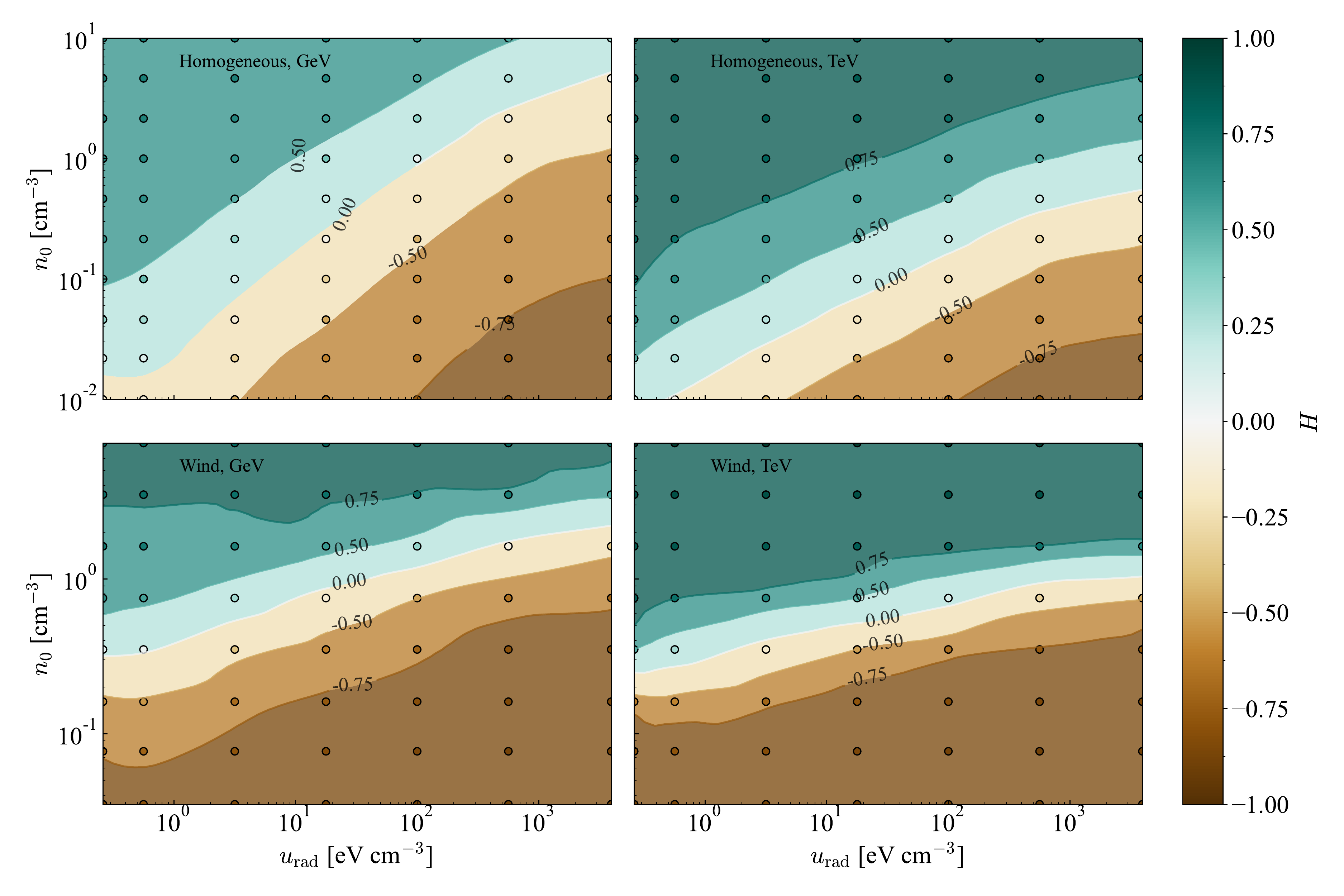}
\caption{Hadronicity as a function of matter density normalization ($n_0$) and radiation energy density ($u_{\rm rad}$). Direct results from our model are presented as data points, while contours in the background are generated by 2D interpolation. The left (right) columns correspond to the GeV (TeV) bands, while the top (bottom) rows correspond to homogeneous (wind) density profiles. Broadly speaking, hadronic emission dominates in environments with high ambient density and low radiation energy density. \label{uni_GeV_e_dens}}
\end{figure*}

To summarize the effect of the environment on an SNR's $\gamma$-ray emission, Figure \ref{uni_GeV_e_dens} shows hadronicity as a function of ambient matter density and radiation energy density. 
At number densities less than order $\sim$0.1--1 cm$^{-3}$, where IC scattering tends to dominate leptonic emission, hadronicity increases linearly with increasing matter density and with decreasing energy density. 
Once the scenario becomes moderately or extremely hadronic, $\pi^0$ decay and relativistic bremsstrahlung become more dominant processes such that the energy density dependence disappears and the hadronicity increases solely with number density. It is also worth noting that hadronic signatures tend to be stronger in the TeV band, largely due to the fact that TeV energies often sample the IC cutoff.

\begin{figure*}[ht]
\includegraphics[width=0.95\textwidth, clip=true,trim= 10 10 10 20]{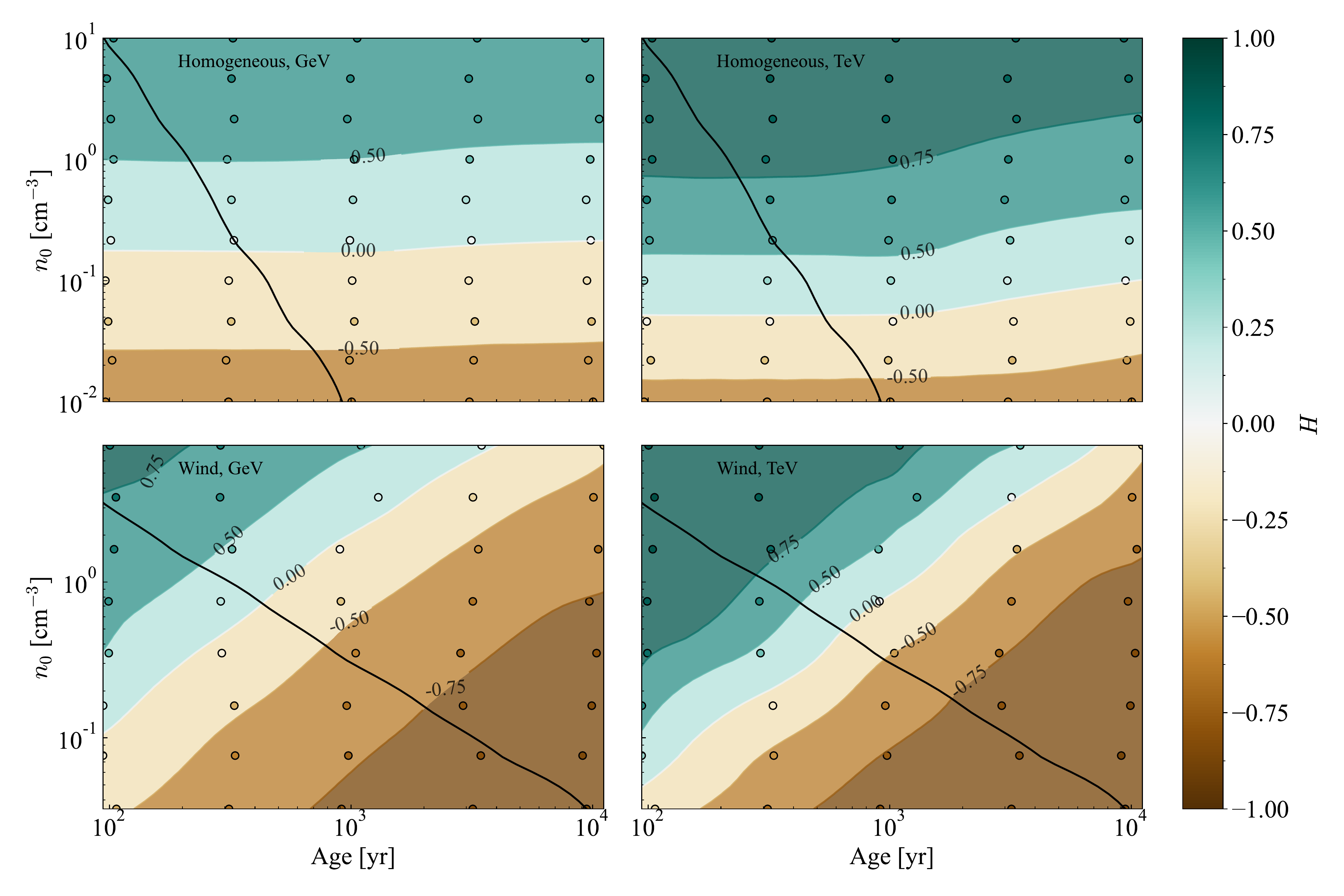}
\caption{Hadronicity as a function of SNR age and ambient density normalization ($n_0$). Direct results from our model are presented as data points, while contours in the background are generated by 2D interpolation. The left (right) columns correspond to the GeV (TeV) bands, while the top (bottom) rows correspond to homogeneous (wind) density profiles. The black lines represent the Sedov-Taylor times for each density profile. In general, young SNRs tend to be the most hadronic. \label{had_age}}
\end{figure*}

We summarize the effect of SNR evolution on hadronicity in Figure \ref{had_age}, which shows hadronicity as a function of SNR age. 
As stated previously, SNRs generally become more leptonic with time. Here we can see that this effect is more pronounced in wind profiles, where the ambient density decreases with radius. 
Homogeneous profiles also exhibit a modest decrease in hadronicity with time in the TeV band, due to the increasing IC cutoff energy; 
in this case, however, we never find situations in which a source transitions from being dominantly hadronic to leptonic.

\begin{figure}[ht]
\includegraphics[width=0.45\textwidth, clip=true,trim= 10 10 10 20]{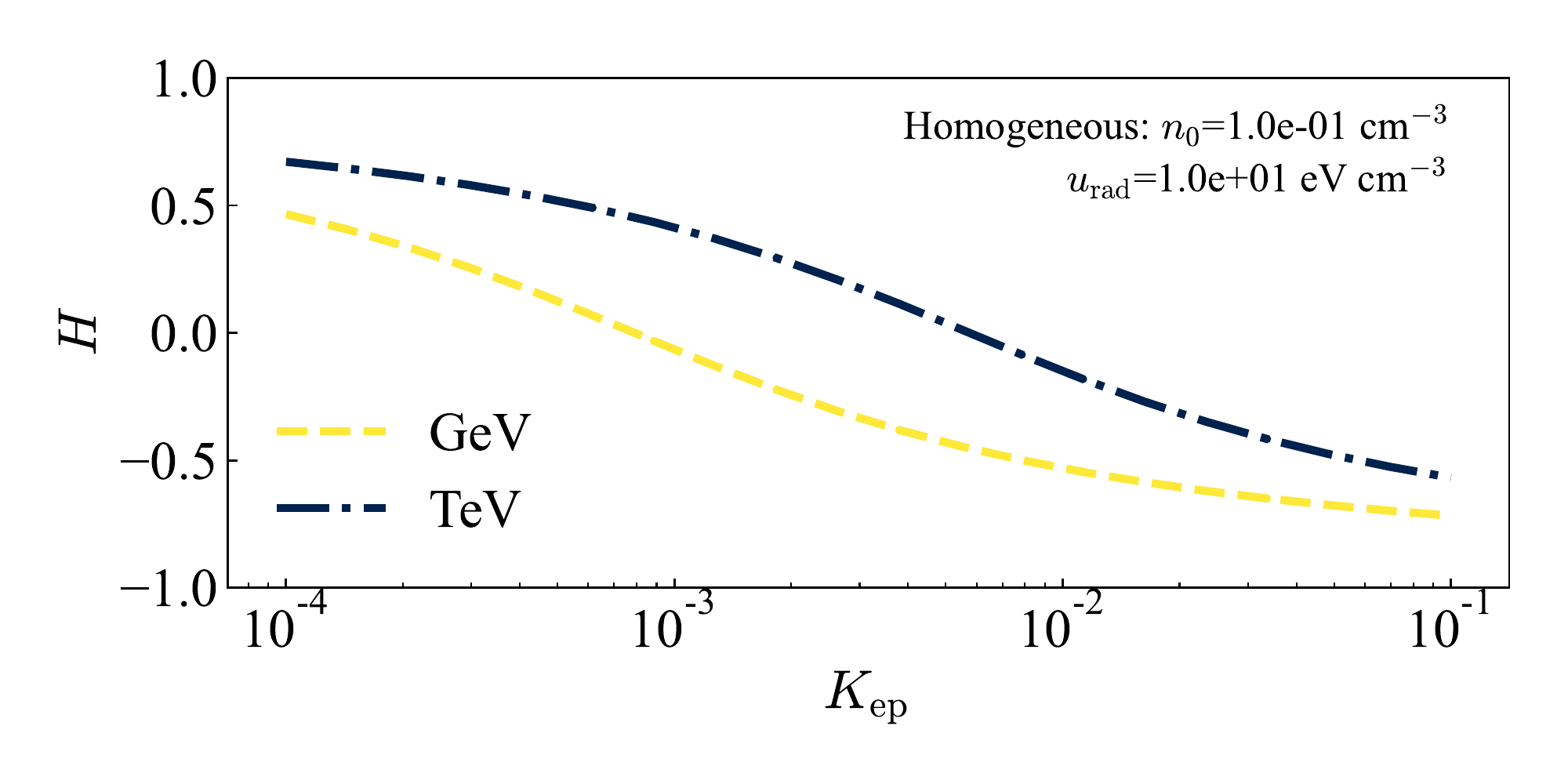}
\caption{Hadronicity in the GeV and TeV bands as a function of the relative normalization of the electron spectrum $K_{\mathrm{ep}}$, for a sample SNR embedded in a low-density environment. \label{kep}}
\end{figure}

Finally, we examine the effect of changing the  normalization of the electron spectrum relative to the ions spectrum, $K_{\rm ep}$.  
Figure \ref{kep} shows the hadronicity of a SNR embedded in a low-density environment as a function of $K_{\mathrm{ep}}$. 
 We allow  $K_{\rm ep}$ to vary from $10^{-2}$ to $10^{-4}$, according to the observational values inferred by the analysis of individual SNRs \citep[e.g.,][]{berezhko+04a,volk+05,morlino+12}, of the radio emission from nearby galaxies \citep[][]{sarbadhicary+17} and kinetic simulations  \citep[e.g.,][]{park+15,xu+20}. 

Both GeV and TeV bands show hadronicity profiles of an arctangent shape, resulting from the direct scaling of the leptonic luminosity with the number of nonthermal electrons. 
The TeV curve, which tends to be more hadronic than its GeV counterpart, appears phase-shifted but otherwise follows the same form. A similar phase shift also occurs when environmental parameters change. 
Thus, when classifying an SNR as ``hadronic'' or ``leptonic,'' marginal cases can be sensitive to uncertainties in $K_{\mathrm{ep}}$
However, in most cases varying $K_{\mathrm{ep}}$ within reasonable values has little impact on the expected nature of the SNR $\gamma$-ray emission.

\section{Discussion} \label{sec:discuss}

Our analysis demonstrates that the apparent hadronicity of an SNR depends strongly on its age and the environment into which it expands. 
Notably, the best SNR candidates for strong hadronic emission are young and expanding into environments with high matter densities and/or low radiation energy densities. 
This conclusion derives from the fact that emission from $\pi^0$ decay scales with matter density, while IC emission scales with radiation energy density. 
Many of the candidates for hadronic SNRs identified in (\cite{caprioli11,acero+15}) indeed are young and/or associated with molecular clouds.
 Likewise, \cite{funk15} collected the spectra of several bright $\gamma$-ray sources, and those that are identified as likely hadronic either are young or exist in high density environments. 

When considering GeV and TeV band spectra, one notable pattern, as mentioned above, is that TeV emission tends to exhibit higher hadronicity. 
This behavior may run counter to expectations since, at TeV energies, the hadronic spectrum is steeper than the leptonic one. However, when measuring hadronicity over a fixed energy band, the dominance of one process over another depends in large part on the position of the high energy cutoff, which, in the case of IC emission, depends on the electron cutoff, which is mediated by synchrotron losses. 


We can summarize the important role of a SNR environment on its $\gamma$-ray emission in terms of a simple scaling relation for hadronicity, assuming IC dominates the leptonic emission. 
Assuming power law particle spectra of the form $ dN_{\rm p}/dE = K_{\mathrm{p}} E^{-q}$ and $dN_{\rm e}/dE = K_{\mathrm{e}} E^{-q}$, we construct expressions for the emissivities of the two radiative processes motivated by the derivations in \cite{longair11} and \cite{ghisellini13}: $\epsilon_{\gamma_{\pi^0}}(E) \propto n K_{\mathrm{p}} E^{-q}$, $\epsilon_{\gamma_{\mathrm{IC}}}(E) \propto u_{\rm rad} K_{\mathrm{e}} E^{-p}$, where $n$ is the ambient proton number density, $u_{\rm rad}$ is the ambient radiation field energy density, and $p=(q-1)/2$. 
Note that these scalings are only valid at energies below the high-energy cutoff for each species; we expect these approximations to break down in the TeV band. 
If we take $K_{\mathrm{ep}} = K_{\mathrm{e}} / K_{\mathrm{p}}$, then the ratio of these two expressions scales as,

\begin{equation}
    \frac{\epsilon_{\gamma_{\pi^0}}}{\epsilon_{\gamma_{\mathrm{IC}}}} \propto \frac{n}{K_{\mathrm{ep}} u_{\rm rad} E^{-(q+1)/2}},
\end{equation}

Using the results of our hadronicity calculations in a homogeneous medium to estimate the normalization of this expression, we obtain,

\begin{equation}
\begin{split}
   &  \frac{\epsilon_{\gamma_{\pi^0}}}{\epsilon_{\gamma_{\mathrm{IC}}}}\simeq 160  \bigg(\frac{10^{-3}}{K_{\rm ep}}\bigg)\bigg(\frac{n}{\text{cm}^{-3}}\bigg)\bigg(\frac{\text{eV cm}^{-3}}{u_{\rm rad}}\bigg)\bigg(\frac{\text{GeV}}{E}\bigg)^{\frac{q+1}{2}}.
\end{split}
\end{equation}

This approximation yields good agreement with the top left panel of Figure \ref{uni_GeV_e_dens}.
The story is somewhat more complicated for expansion in a wind profile, for which this simple, single-zone model is not necessarily a good approximation.
Also, it is possible that the IC emission in the TeV band is overestimated because synchrotron losses may steepen the electron spectrum with respect to the ions' \citep[][]{diesing+19}.
However, in the absence of detailed information about a SNR expansion history, this expression still holds as a rough estimate of the hadronicity.

Generally speaking, the overall trends identified are consistent across both homogeneous and wind profiles, with the most notable difference being that in a decreasing density profile, SNRs may exhibit a more and more leptonic emission as they age. 
Physically speaking, any wind profile should terminate at some finite distance \citep[][]{weaver+77,ptuskin+05}, beyond which the emission should be similar to the one of a SNR expanding in the homogeneous ISM \citep[][]{caprioli11}.
Furthermore, beyond a power-law scaling with radius, none of our profiles include inhomogeneities (clumps, molecular clouds, ISM-scale gradients), which would certainly be present in realistic scenarios. Since the matter density of clouds would be greater than their ambient surroundings, our results suggest that that they would increase the hadronicity of an observed source.

\section{Conclusion} \label{sec:conclusion}

In summary, we modeled time-dependent, multi-zone, CR acceleration in an evolving SNR using a semi-analytical implementation of non-linear DSA;
the goal is to understand the factors that influence whether a SNR emission is dominated by  hadronic or leptonic processes. 
We find that, for a fixed supernova explosion, the dominance of hadronic or leptonic emission is governed by environmental factors, rather than by the underlying nature of particle acceleration. 

Furthermore, we find that SNRs tend to appear more leptonic as they age--particularly in the TeV band--due to decreases in the amplified magnetic field and thus increases in the  synchrotron-modulated maximum electron energy.
This transition is even more pronounced in SNRs expanding into media with decreasing density (i.e., stellar winds).
Thus, our findings suggest that the best candidates bearing signatures of hadron acceleration are young, core-collapse SNRs, as well as SNRs interacting with molecular clouds.

More quantitatively, SNRs expanding into media with $[n/(\text{cm}^{-3})]/[u_{\rm rad}/(\text{eV cm}^{-3})] \gtrsim 3$ are likely to exhibit hadronic signatures even in the case of very efficient electron acceleration ($K_{\rm ep} \gtrsim 10^{-2}$). 
These findings may guide the missions of very-high energy $\gamma$-ray observatories such as H.E.S.S., MAGIC, VERITAS, LHAASO, and, in the near future, CTA.

\begin{acknowledgments}
This research was partially supported by NASA grant 80NSSC20K1273 and the NSF grants AST-1909778, AST-2009326 and PHY-2010240.
\end{acknowledgments}

\bibliography{Total}
\bibliographystyle{aasjournal}

\end{document}